\documentclass[useAMS,usenatbib]{mnras}
\usepackage{graphicx}

\title[On the Hoffmeister family]
{The Hoffmeister asteroid family}
\author[V. Carruba, B. Novakovi\'{c}, S. Aljbaae]{V. Carruba$^{1}$\thanks{E-mail: vcarruba@feg.unesp.br}, B.Novakovi\'{c}$^{2}$, S. Aljbaae$^{1}$\\
$^{1}$UNESP, Univ. Estadual Paulista, Grupo de din\^{a}mica Orbital e
  Planetologia, Guaratinguet\'{a}, SP, 12516-410, Brazil.\\
  $^{2}$University of Belgrade, Department of Astronomy,
  Faculty of Mathematics, 11000 Belgrade, Serbia.\\
}

\begin{document}

\date{Accepted 2016 November 18. Received 2016 November 17; in original form 2016 September 21.}

\pagerange{\pageref{firstpage}--\pageref{lastpage}} \pubyear{2016}

\maketitle

\label{firstpage}

\begin{abstract}
The Hoffmeister family is a C-type group located in the central main belt.
Dynamically, it is important because of its interaction with the ${\nu}_{1C}$
nodal secular resonance with Ceres, that significantly increases the
dispersion in inclination of family members at lower semi-major axis.
As an effect, the distribution of inclination values of the Hoffmeister
family at semi-major axis lower than its center is significantly leptokurtic,
and this can be used to set constraints on the terminal ejection velocity
field of the family at the time it was produced.  By performing
an analysis of the time behaviour of the kurtosis of the $v_W$ component
of the ejection velocity field (${\gamma}_2(v_W)$), as obtained from
Gauss' equations, for different fictitious Hoffmeister families with
different values of the ejection velocity field, we were able to exclude
that the Hoffmeister family should be older than 335 Myr.  Constraints
from the currently observed inclination distribution of the Hoffmeister
family suggest that its terminal ejection velocity parameter $V_{EJ}$
should be lower than 25~m/s. Results of a Yarko-YORP Monte Carlo method
to family dating, combined with other constraints from inclinations and 
${\gamma}_2(v_W)$, indicate that the Hoffmeister family should be
$220^{+60}_{-40}$~Myr old, with an ejection parameter $V_{EJ}= 20\pm5$~m/s.
\end{abstract}

%

\section{Introduction}
\label{sec: intro}

The Hoffmeister family was identified in \citet{Milani_2014}, and
more recently in \citet{Nesvorny_2015}  with a Family Identification
Number (FIN) equal to 519. As originally observed by Novaković et al. (2015),
this family is characterized by its interaction with
the ${\nu}_{1C}$ nodal secular resonance with Ceres, whose effect is
to significantly spread the distribution in inclination of family members
for semi-major axis lower than $\simeq 2.78$~au.  \citet{Carruba_2016}
also identified this family as one of the eight groups
characterized by having the most leptokurtic
distribution of the $v_W$ component of terminal ejection velocities field,
that is closely related to the inclination through the third Gauss
equation (see, for instance, Eq.~3 in \citet{Carruba_2016}).
While most families are formed with an originally leptokurtic distribution
of ejection velocities (see \citet{Carruba_2016} for an explanation
of the role that escape velocities from parent body have in creating
originally leptokurtic distributions of terminal ejection velocities),
in the absence of dynamical mechanisms able to change the inclination
of family members, the distribution of $v_W$ tends in time to become more
mesokurtic, or Gaussian.  This, however, is not the case
for the Hoffmeister family, whose interaction with the ${\nu}_{1C}$ nodal
secular resonance significantly shaped its inclination distribution.

The peculiar nature of the Hoffmeister family allows for the use
of techniques of family dating not available for other asteroid groups.
In particular, the study of the time behavior of the kurtosis of the
$v_W$ component of the ejection velocity field (${\gamma}_2(v_W)$),
as performed by \citet{Carruba_2016b, Carruba_2016c} for the Astrid, Gallia,
Barcelona, and Hansa families, could provide invaluable constraints
on the family age and on the $V_{EJ}$ parameter describing the
standard deviation of the initial ejection velocity field, assumed
as Gaussian.  By observing by what time the current value of ${\gamma}_2(v_W)$
is reached, for fictitious Hoffmeister families with different values
of $V_{EJ}$, and by imposing constraints on the most likely value
of this parameter, based on the current inclination distribution 
of the part of the Hoffmeister family not affected by the ${\nu}_{1C}$
secular resonance, we can then obtain constraints on both the most likely
values of family age and $V_{EJ}$, not available for regular families.

\section{Family identification and dynamical properties}
\label{sec: fam_ide}

As a first step in our analysis, we selected the (1726) Hoffmeister
family\footnote{For the sake of brevity, the identification number
  of families discussed in this paper will only be provided once.  After
  that, families will be referred to only by the parent body name.}, 
as identified in \citet{Nesvorny_2015} using the Hierarchical Clustering Method
(HCM, \citep{Bendjoya_2002}) and a cutoff of 45~m/s.  1819 members of the
Hoffmeister dynamical group were identified in that work.
\citet{Milani_2014} identified 1905 members of the Hoffmeister
family, with an $(a,e)$ and $(a,\sin{(i)})$ distribution very similar to
that observed for the \citet{Nesvorny_2015} group.  For the sake of
consistency, in this work we will use the data from \citet{Nesvorny_2015},
but we will discuss any significant difference with results from
\citet{Milani_2014}, when appropriate.  Using the criteria defined in
\citet{Carruba_2016} we defined as objects in the
local background of the Hoffmeister family as those with 
synthetic proper elements\footnote{We use data from the AstDyS site 
(http://hamilton.dm.unipi.it/astdys, \citet{Knezevic_2003}), accessed on July
3$^{rd}$, 2016.}, whose values of proper $e$ and $\sin{(i)}$ are 
in a range from the family barycenter to within
four standard deviations of the observed distribution for the Hoffmeister
family, namely from 0.0300 to 0.071 in proper $e$ and from 0.054
to 0.102 in $\sin{(i)}$.  The minimum value of proper $a$ was
chosen from the minimum value of Hoffmeister members minus 0.02 au,
the averaged expected orbital mobility caused by close encounters with
massive asteroids over 4 Byr \citep{Carruba_2013}. The maximum $a$ value
was taken at the center of the 5J:-2A mean-motion
resonance.  Namely, this corresponds to an interval in $a$ from 2.730 au to
2.825 au.  Overall, we found 4781 asteroids in the local background 
of the Hoffmeister family so defined, also including known members of
other families.

\citet{Carruba_2009}, in his analysis of the (363) Padua family found that
the main asteroid families in the local background of the Hoffmeister cluster 
are the (847) Agnia and Padua groups, both characterized by their
interaction with the $z_1$ secular resonance \citep{Vokrouhlicky_2006b,
  Carruba_2009}. This analysis was essentially confirmed by
\citet{Nesvorny_2015}.  Apart from the cited Agnia and Padua families, these
authors also identified the (2732) Witt family at higher inclinations.
Other families in the region, such as the groups of (128) Nemesis,
(1668) Hanna, and (1222) Tina, do not have family
members in the local background of Hoffmeister defined according
to our criteria.   Concerning the two major other families in
the region, Padua and Agnia, despite some difference in terms of family
membership between the Nesvorn\'{y} and Milani groups for the Padua family,
that has 1087 and 864 members in these two classifications respectively,
the corresponding distributions in the $(a, e)$ and $(a, \sin{(i)})$
planes are similar.  The only significant difference
in the distribution in the $(a,\sin{(i)})$ plane for the two Padua families
is that the Nesvorn\'{y} group has a small population of objects at
$a > 2.78$~au not visible in the Milani group.  The case of the Agnia
family is a bit more problematic.  This family has 3033
members in Milani et al. (2014), and only 2125 members in Nesvorn\'{y} et al.
(2015).  Most of the difference in family memberships is caused
by the presence of multi-opposition objects in the Milani group, that
are not accounted for in the Nesvorn\'{y} family.  Also, the Milani
Agnia family extends beyond the 3-1-1 three-body resonance, while the
Nesvorn\'{y} one does not.  Quite interestingly, \citet{Milani_2014}
and \citet{Spoto_2015} identify the (3395) Jitka sub-family inside
the Agnia family, that is quite distinguished in term of physical
properties from the rest of the Agnia family, and that will be
further discussed in the next section.

These differences in family memberships for the Agnia and, in a lesser
measure, the Padua groups can a bit affect the determination of the
local background, but should not change our overall conclusions. Therefore,
since there is a good agreement between the two Hoffmeister families,
in terms of family membership and other properties, in this work we will
use the families as determined in \citet{Nesvorny_2015}.

\begin{figure*}
  \centering
  \begin{minipage}[c]{0.49\textwidth}
    \centering \includegraphics[width=3.1in]{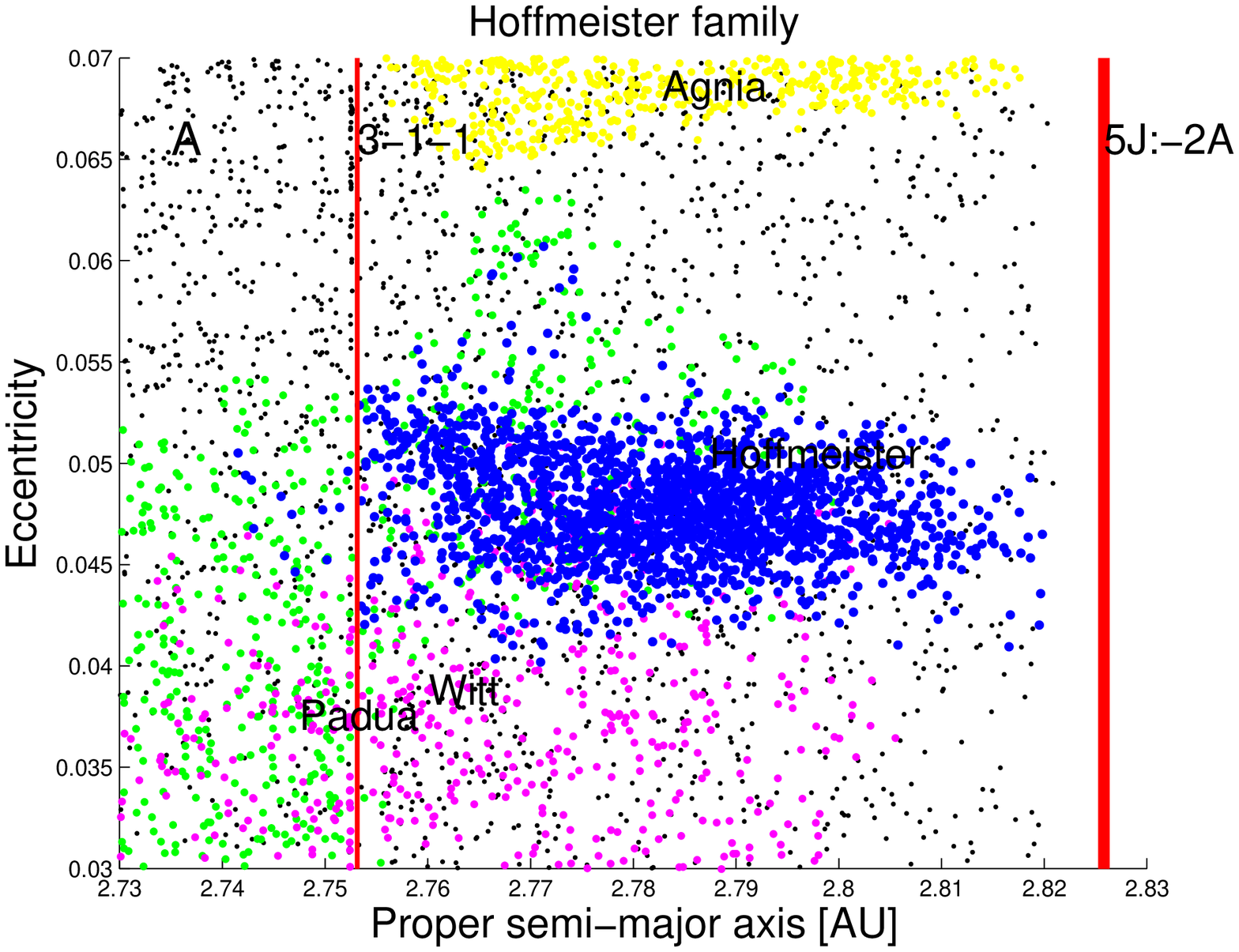}
  \end{minipage}%
  \begin{minipage}[c]{0.49\textwidth}
    \centering \includegraphics[width=3.1in]{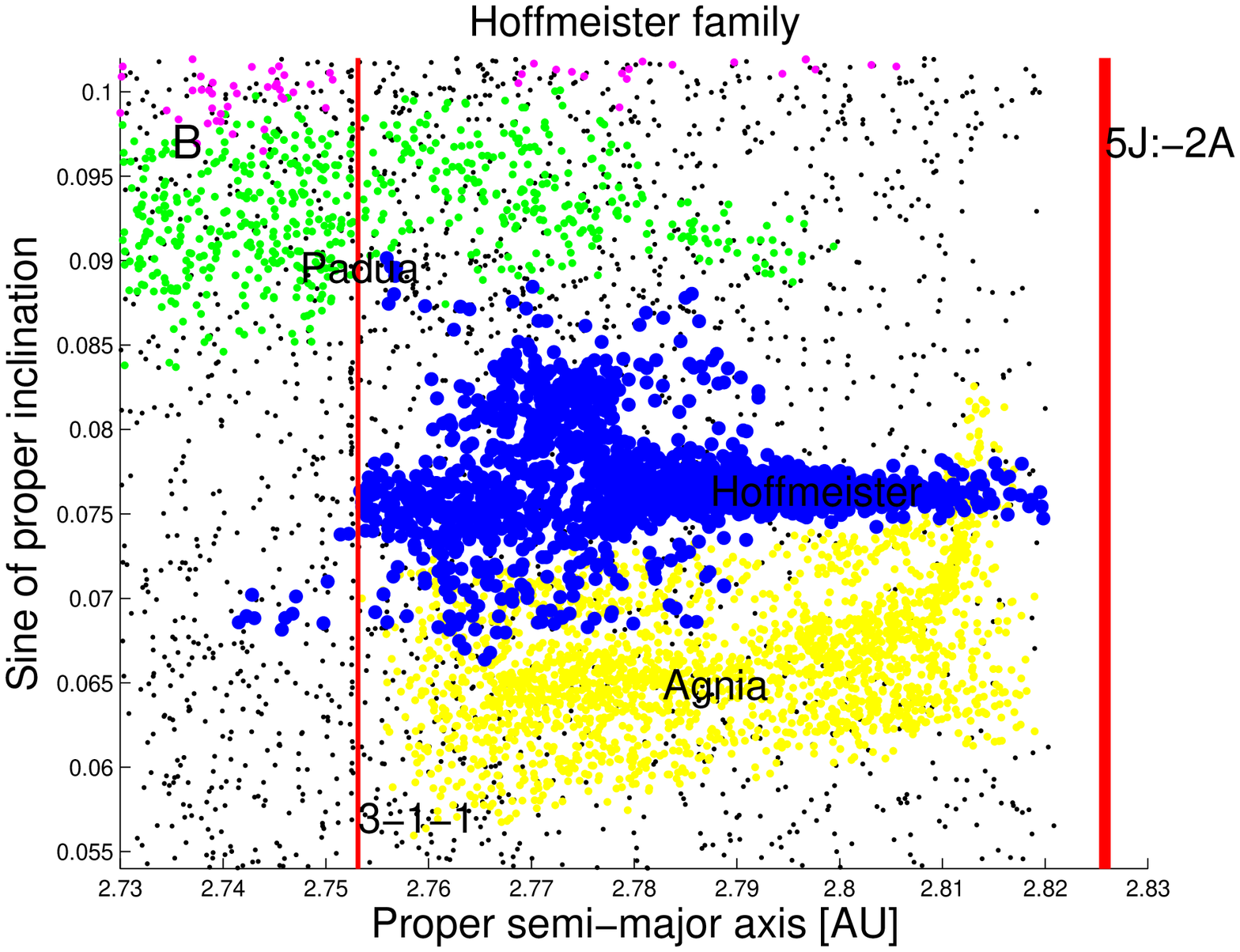}
  \end{minipage}
\caption{An $(a,e)$ (panel A) and $(a,\sin{(i)})$ (panel B) 
projection of asteroids
in the local background of the Hoffmeister family.  Vertical lines display
the location of the main mean-motion resonances in the region.  Blue full
dots show the orbital location of members of the Hoffmeister family, 
green dots those of the Padua family, yellow dots those of the Agnia family,
and magenta dots are associated with the orbits of members of the Witt family.
Black dots show the location of asteroids in the Hoffmeister local background.}
\label{fig: Back_hoff}
\end{figure*}

After removing members of the Padua, Agnia, and Witt groups, the local
background consisted of 1721 objects.  Fig.~\ref{fig: Back_hoff} display
a projection as black dots of these asteroids in the $(a,e)$ (panel A)
and $(a,\sin{(i)})$ plane.  Members of the Hoffmeister, Padua, Agnia, and Witt
families are shown as full blue, green, yellow, and magenta dots.
As originally observed in \citet{Novakovic_2015}, the left side of the 
Hoffmeister family is quite more spread in inclination than its right
side, because of the interaction of this family with the ${\nu}_{1C}=s-s_C$
linear secular resonance with Ceres.  To start assessing the 
dynamical importance of this resonance, we obtained a dynamical map
of synthetic proper elements, with the method discussed in 
\citet{Carruba_2010}, for particles subjected under the gravitational 
influence of all planets, plus Ceres, Pallas, and Vesta\footnote{Dynamical
maps in the $(a,e)$ and $(a,\sin{(i)})$ for the cases without massive
asteroids were obtained for the region of the Padua family in 
\citet{Carruba_2009}.  Interested readers could find more information
about these results in that paper.}.   We integrated 3000 particles
in a 50 by 60 grid in osculating initial $(a,\sin{(i)})$ plane,
with a step of 0.02 au in $a$ and 0.1 degrees in $i$.  Initial values
of $a$ and $i$ were 2.73 au and 0 degrees, respectively.  The eccentricity
and the other angles of the test particles were those of (1726) Hoffmeister
at J2000.

\begin{figure*}
  \centering
  \begin{minipage}[c]{0.49\textwidth}
    \centering \includegraphics[width=3.1in]{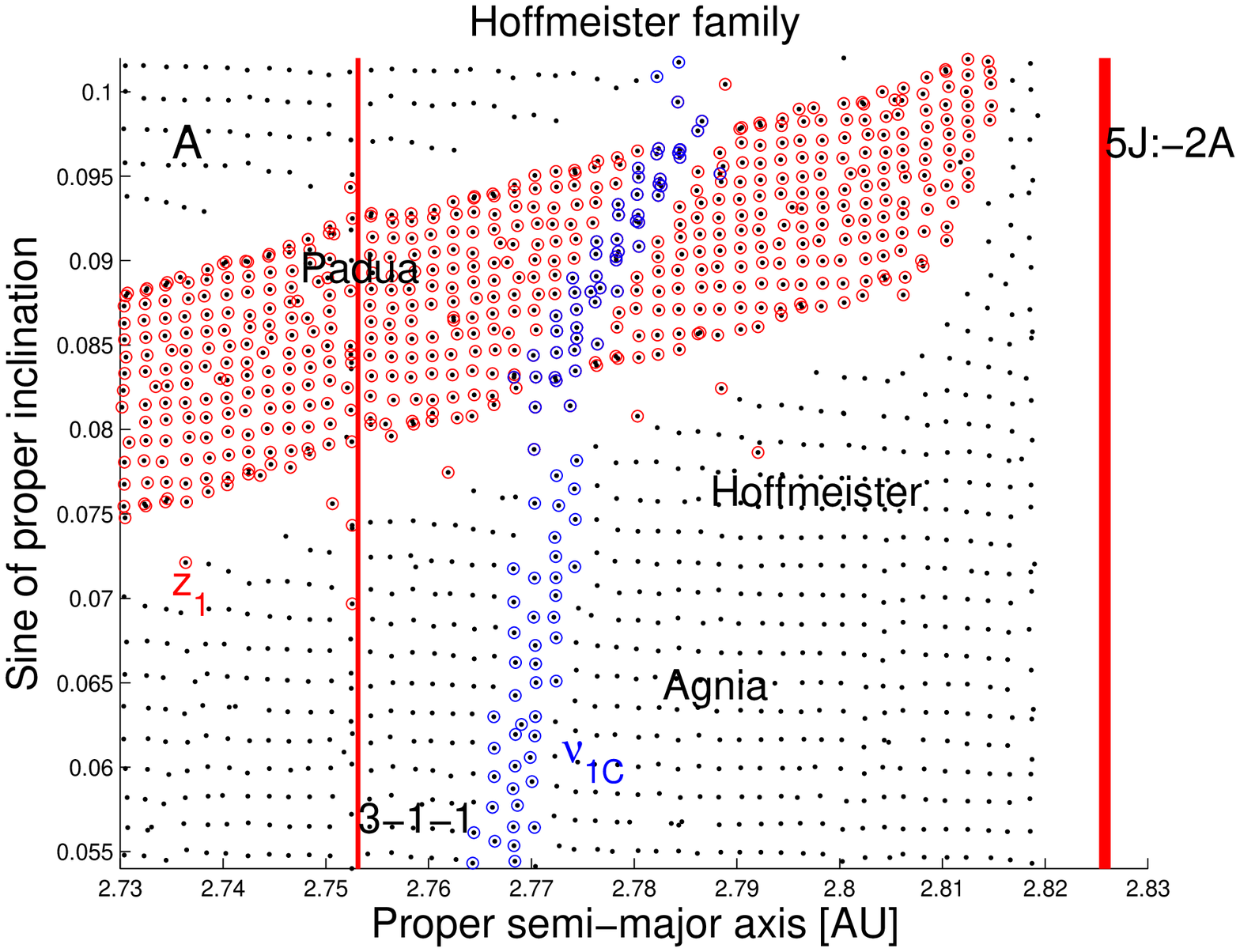}
  \end{minipage}%
  \begin{minipage}[c]{0.49\textwidth}
    \centering \includegraphics[width=3.1in]{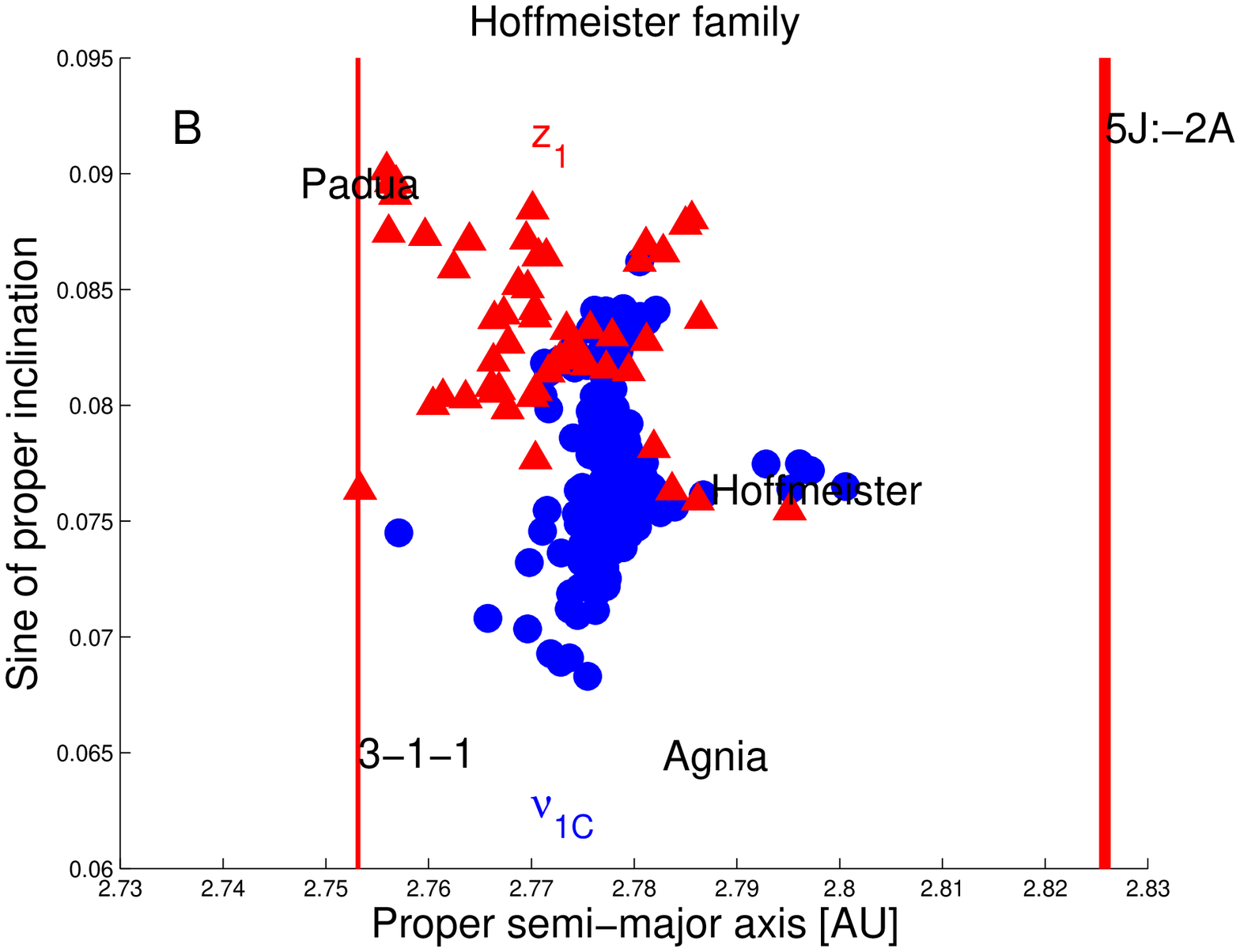}
  \end{minipage}
\caption{A dynamical map in the proper $(a,\sin{(i)})$ domain for the orbital
region of the Hoffmeister family, considering
the effect of Ceres, Vesta and Pallas as massive perturbers (panel A).  Black
dots identify the values of proper $a$ and $\sin{(i)}$ for the 
integrated test particles.  Blue and red full dots are associated
with likely resonators of the ${\nu}_{1C}$ and $z_1$ secular resonance,
respectively. Other symbols are the same as in Fig.~\ref{fig: Back_hoff}.
Panel B display an $(a,\sin{(i)})$ projection of Hoffmeister members
in ${\nu}_{1C}$ (blue full dots) and $z_1$ (red full triangles) librating 
states, respectively.}
\label{fig: map_ai_cpv}
\end{figure*}

Our results are shown in Fig.~\ref{fig: map_ai_cpv}, panel A.  Values of the 
proper $(a,\sin{(i)})$ of each particle are shown as black dots.
Blue and red full circles are associated
with likely resonators of the ${\nu}_{1C}$ and $z_1$ secular resonance,
defined as objects whose $s$ and $g+s$ frequencies are to within
$\pm 0.2$ arcsec/yr from the proper node frequency of Ceres for the
${\nu}_{1C} =s -s_C$ resonance ($s_c = -59.17$ arcsec/yr, 
\citep{Knezevic_2003}), and from the sum of the proper pericenter
and node frequencies of Saturn for the $z_1 = g+s-g_6-s_6$ resonance
($g_6+s_6 = 1.90$ arcsec/yr \citep{Knezevic_2003}).  As discussed
in \citet{Carruba_2009}, the $z_1$ resonance plays a major role
in the dynamical evolution of the Padua family, while the ${\nu}_{1C}$
has a pivotal role in the evolution of the Hoffmeister group
\citep{Novakovic_2015}.   To check for the effects of these
two resonance on the dynamical evolution of the Hoffmeister family,
we also integrated 1819 members of the family and checked the
time behavior of the ${\nu}_{1C}$ and $z_1$ resonant arguments over 20 Myr.
We found 180 and 54 members of the family in librating states of these
two resonances, respectively.  Their $(a,\sin{(i)})$ projection 
is shown in Fig.~\ref{fig: map_ai_cpv}, panel B.  Hoffmeister members
are dispersed in inclination after interacting with the ${\nu}_{1C}$
resonance, as originally observed by \citet{Novakovic_2015}.  A significant
fraction of Hoffmeister members (about 3\%) reached values of inclination 
large enough to allow for capture into the $z_1$ resonance.  We expect
that some of the past members of the Hoffmeister family could have been
drawn to the region of the Padua family and be current interlopers
of that group.

\section{Physical properties}
\label{sec: Taxonomy}

\begin{figure*}
  \centering
  \begin{minipage}[c]{0.49\textwidth}
    \centering \includegraphics[width=3.1in]{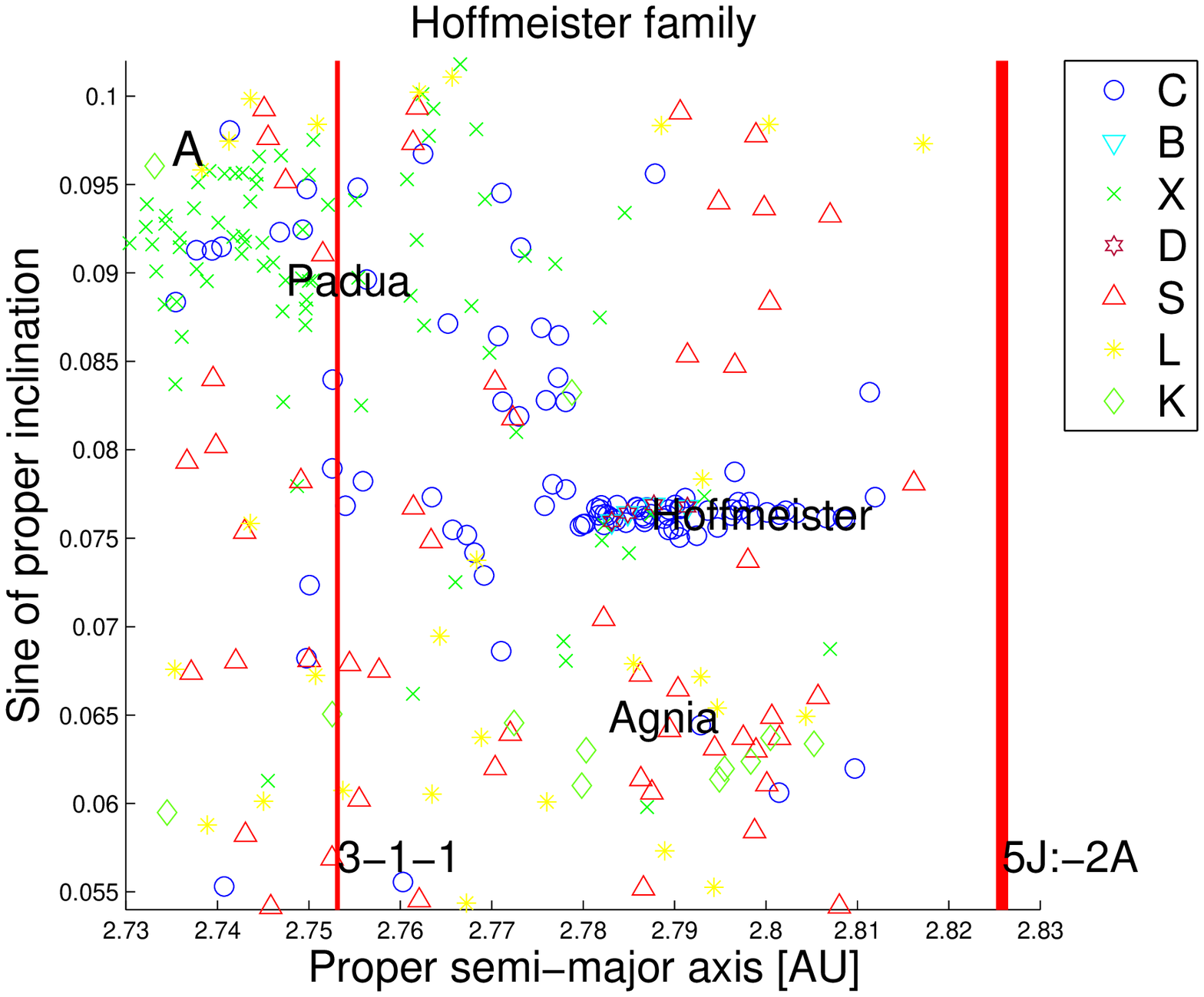}
  \end{minipage}%
  \begin{minipage}[c]{0.49\textwidth}
    \centering \includegraphics[width=3.1in]{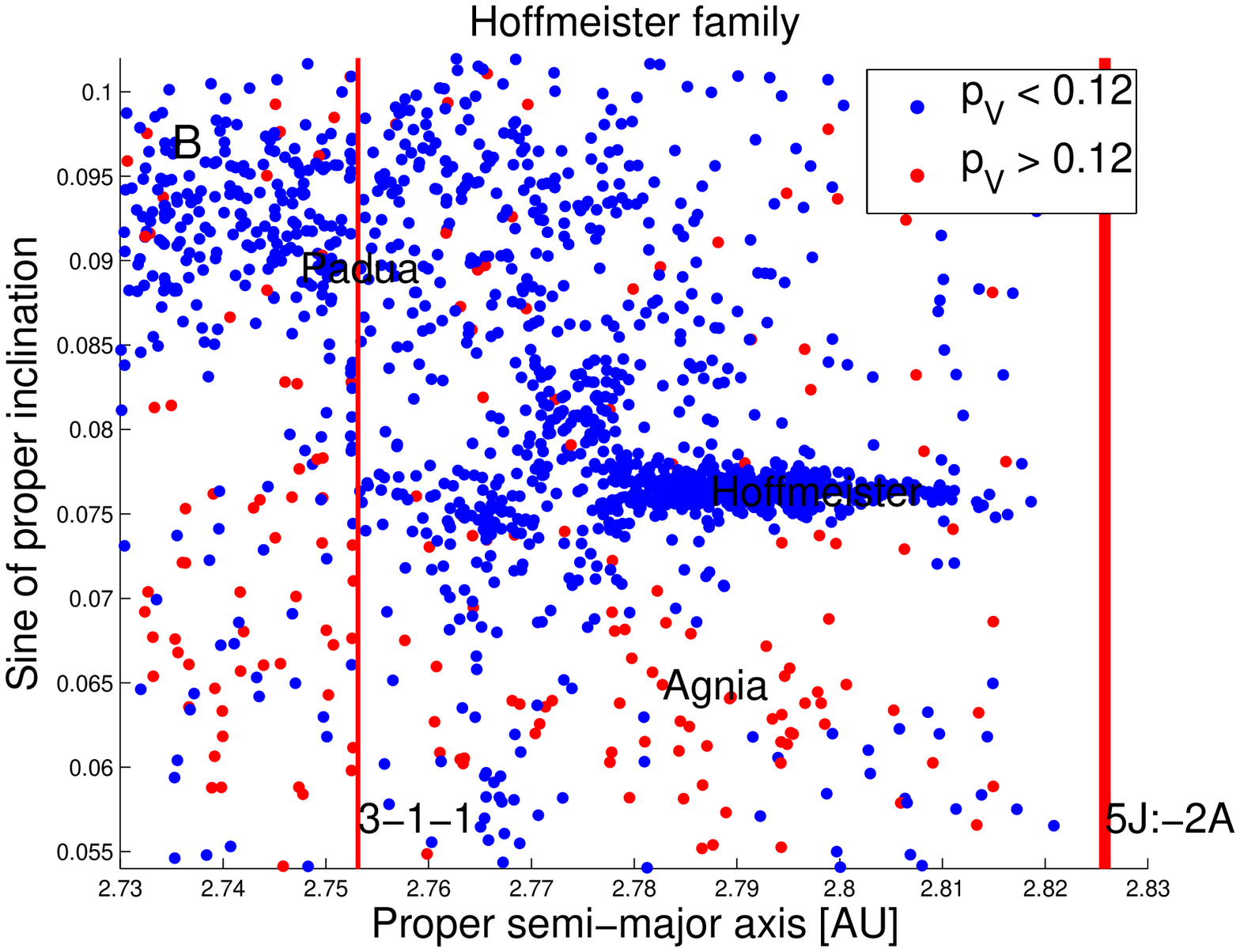}
  \end{minipage}
\caption{An $(a,\sin{(i)})$ projection of asteroids with SDSS-MOC4 
taxonomic classification, according to the \citet{DeMeo_2013} method.
(panel A).  Symbols identify the different asteroids according to
the scheme outlined in the figure legend.  Panel B displays
an $(a,\sin{(i)})$ projection of asteroids with WISE albedo $p_V$ lower
than 0.12 (full blue dots), and higher than 0.12 (full red dots).}
\label{fig: Backround_SDSS_WISE}
\end{figure*}

After revising the effect of the local dynamics, we now turn our attention
to the taxonomic properties of local objects.  \citet{Carruba_2009} discuss
properties of asteroids in the orbital proximity of the Padua family,
interested readers could find more information in that paper.  Concerning
objects in the background of the Hoffmeister family, as defined
in this section, we found 287 objects with photometric data 
in the Sloan Digital Sky Survey-Moving Object Catalog data, 
fourth release (SDSS-MOC4 hereafter, \citep{Ivezic_2001}) in thi s region,
71 of which are members of the Hoffmeister dynamical family.
If we consider all available data, regardless of its error,
1515 objects have geometric albedo and absolute magnitude information 
in the WISE and NEOWISE database \citep{Masiero_2012}. 
Fig.~\ref{fig: Backround_SDSS_WISE} displays asteroids 
with their classification obtain with the 
method of \citet{DeMeo_2013} (panel A).  Panel B
displays objects with WISE geometric albedo $p_V$ with values compatible
with a C-complex taxonomy ($p_V < 0.12$, blue full dots) and 
with an S-complex taxonomy, ($0.12 < p_V < 0.30$, red full dots,
\citet{Masiero_2012}.  As can be seen from the figure, the Hoffmeister
family is compatible with a C-type composition, the Padua family was
probably originated from the break-up of an X-type asteroid, and most of
the members of the Agnia family are S-type, which also confirms the analysis
of \citet{Nesvorny_2015}.  As discussed in \citet{Spoto_2015}, the
Jitka sub-family in the Agnia family is characterized by higher
albedo values than that of the rest of the family.   While most
of Agnia members have $p_V = 0.15\pm 0.01$, members of the Jitka
sub-family have $p_V = 0.31\pm 0.04$. The Witt family, not visible in
the figure, is a S-type group.  The mean albedo value for the Hoffmeister,
Padua, and Agnia family is of $(0.05 \pm 0.025), (0.07\pm0.03)$,
and $(0.15\pm0.01)$, respectively.

\begin{figure*}
  \centering
  \begin{minipage}[c]{0.49\textwidth}
    \centering \includegraphics[width=3.1in]{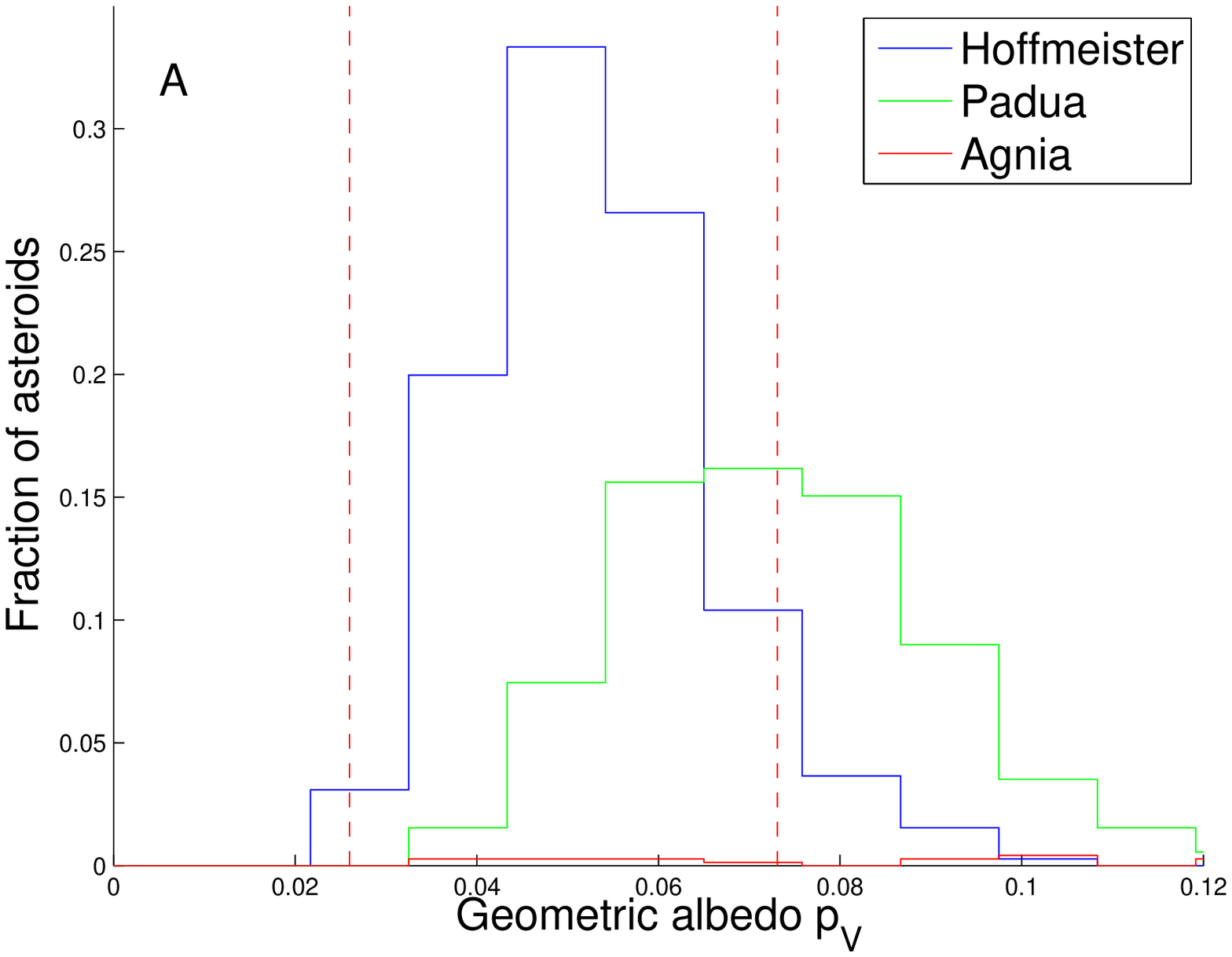}
  \end{minipage}%
  \begin{minipage}[c]{0.49\textwidth}
    \centering \includegraphics[width=3.1in]{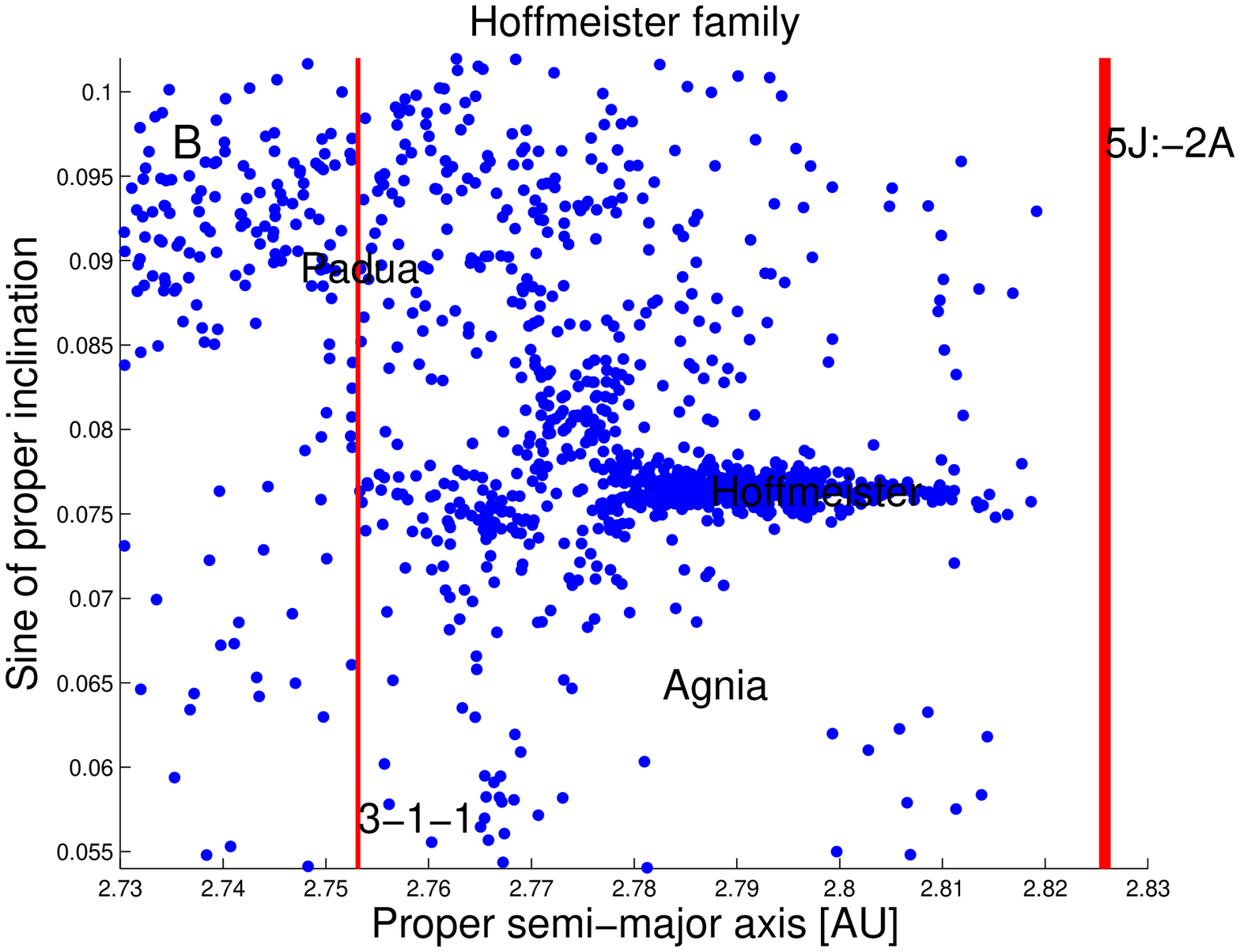}
  \end{minipage}
\caption{An histogram of the WISE albedo
data for the asteroids members of the Hoffmeister (blue line),
Padua (green line) and Agnia (red line) families (panel A). Panel B
displays an $(a,\sin{(i)})$ projection of objects with WISE albedo
in the 1-sigma interval associated with the Hoffmeister family.}
\label{fig: Wise_halo}
\end{figure*}

Concerning halo objects, in the local background of the Hoffmeister
family only the Hoffmeister family itself is associated with a C-type
composition.  It is reasonable therefore to conclude that the 95 
asteroids with a SDSS-MOC4 C-type compatible composition in the region
(see Fig.~\ref{fig: Backround_SDSS_WISE} could all be potentially associated 
with the Hoffmeister family.  C-type objects in the region of the Padua
family could be potential former members of the Hoffmeister family
that reached this region with the mechanism of dynamical diffusion
(capture in the ${\nu}_{1C}$ resonance and then in the $z_1$) discussed
in the previous section.  If we consider the albedo data, 
Fig.~\ref{fig: Wise_halo}, panel A, displays an histogram of the WISE albedo
data for the asteroids members of the Hoffmeister (blue line),
Padua (green line) and Agnia (red line) families.  Vertical red 
dashed lines show the 1-sigma range of albedo values for the
Hoffmeister family.  There is a region of overlapping for asteroids
belonging to the Hoffmeister and Padua family, (at 1-sigma level the
Padua family covers the range of $p_V = 0.07\pm0.03)$,
so that it is not possible to distinguish objects originating from one of these
two families only based on albedo (the contribution from
the Agnia family in this 1-sigma interval is essentially negligible).  
If we limit our data to asteroid in the 
1-sigma $p_V$ interval ($0.025 < p_V < 0.075$, 
Fig.~\ref{fig: Wise_halo}, panel B), then we can safely
conclude that objects at inclination lower than that of the Hoffmeister
barycenter should be most likely originate from this family.  However, 
no conclusion could be safely reached for objects at higher inclinations.

\section{Constraints on terminal ejection velocities from the current
inclination distribution}
\label{sec: inc_constr}

\citet{Nesvorny_2015} estimate the age of the Hoffmeister
family to be $210\pm10$ Myr, while \citet{Spoto_2015}, using a V-shape 
criteria, evaluate the family to be $330\pm100$ Myr old,
and found no asymmetry between the left and right side of the family
semi-major axis distribution, concerning the estimation of the 
group age.  As discussed in \citet{Carruba_2016b} Monte Carlo methods 
\citep{Vokrouhlicky_2006a, 
Vokrouhlicky_2006b, Vokrouhlicky_2006c} that simulates the evolution 
of the family caused by the Yarkovsky and YORP effects, where
YORP stands for Yarkovsky-O'Keefe-Radzievskii-Paddack effect, could also be used
to obtain estimates of the age and terminal ejection velocities of the 
family members (these models will be referred as ``Yarko-Yorp'' models 
hereafter).

However, the age estimates from these methods depend on 
key parameters describing the strength of the Yarkovsky force, such as the 
thermal conductivity $K$ and bulk and surface density ${\rho}_{bulk}$ and 
${\rho}_{surf}$, that are in many cases poorly known.
Before attempting our own estimate of the family age and terminal 
ejection velocity field, here we analyze what constraints could be obtained
on the possible values of terminal ejection velocities of the 
original Hoffmeister family from its current inclination distribution. 

\begin{figure*}
  \centering
  \begin{minipage}[c]{0.49\textwidth}
    \centering \includegraphics[width=3.1in]{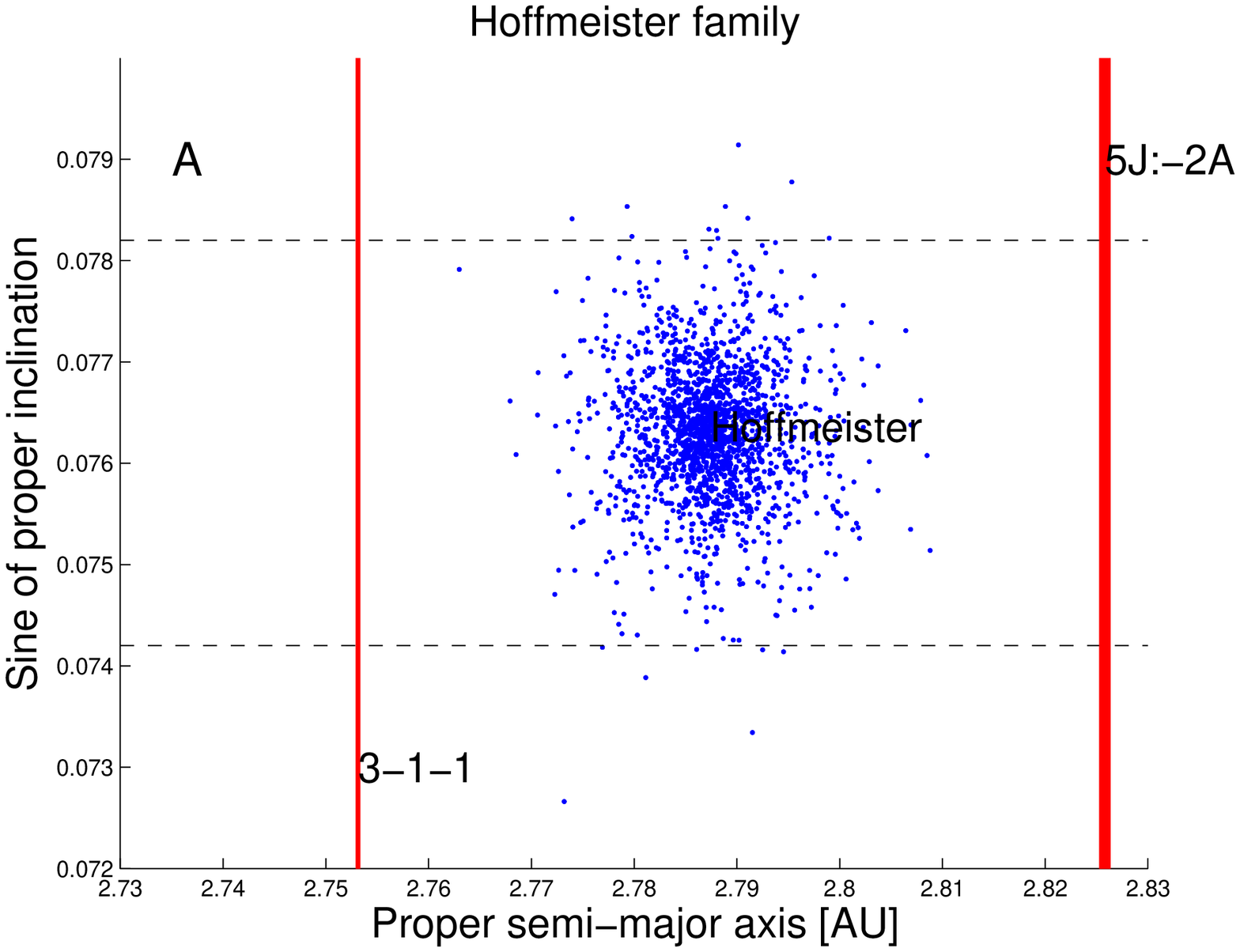}
  \end{minipage}%
  \begin{minipage}[c]{0.49\textwidth}
    \centering \includegraphics[width=3.1in]{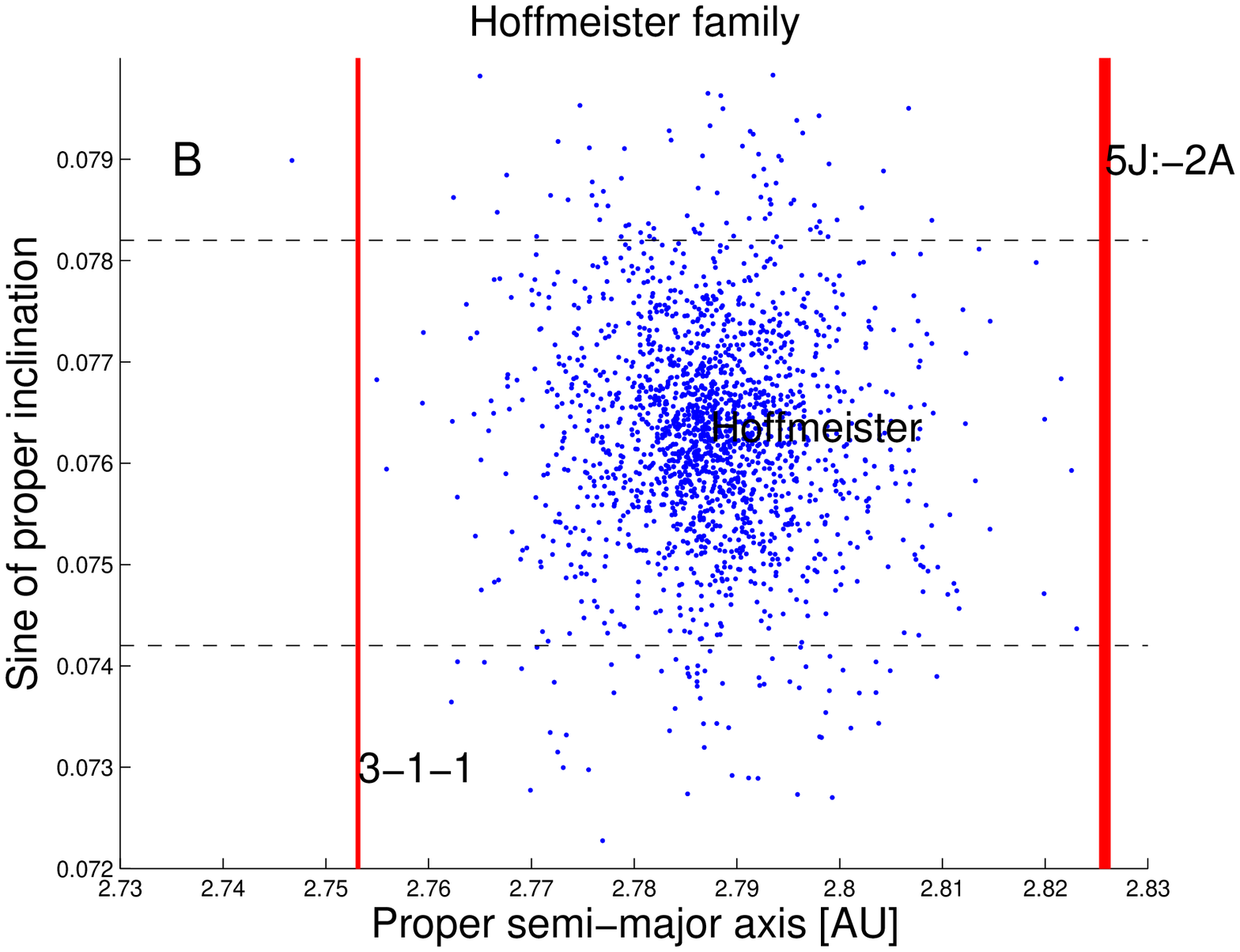}
  \end{minipage}
\caption{An $(a,\sin{(i)}$ projection of the initial orbital dispersion
of a family generated with $V_{EJ} = 15$~m/s (panel A) and $V_{EJ} = 25$~m/s 
(panel B). The dashed lines show the minimum and maximum values of 
$\sin{(i)}$ currently observed for members of the Hoffmeister family with 
$a > 2.795$~au, i.e., those that did not yet interacted with the $s-s_C$ 
secular resonance. The other symbols have the same meaning as in 
Fig.~\ref{fig: Back_hoff}.}
\label{fig: inc_constr}
\end{figure*}

Assuming, in first approximation, that the original ejection velocity
field of the Hoffmeister family could be approximated as isotropic
(see \citet{Carruba_2016} for a discussion of the caveats on this 
hypothesis), we can model the distribution of ejection velocities
of asteroids with a Gaussian distribution  
whose standard deviation follows the relationship:

\begin{equation}
V_{SD}=V_{EJ}\cdot(5km/D),
\label{eq: V_EJ}
\end{equation}

\noindent
where $V_{EJ}$ is the terminal ejection velocity parameter to be estimated,
and $D$ is the asteroid diameter.  \citet{Broz_2013} estimated that 
the ratio of the radius of the parent body with that of the largest
fragment was of 0.14, which yields an estimate of 90.85 km for the 
parent body diameter, and an escape velocity of 39.6~m/s, assuming
a mean density for the parent body of 1300~kg/m$^3$, typical
of C-type asteroids.  If we only 
consider objects with $a > 2.795$~au, so as to eliminate the asteroids 
that interacted with the $s-s_C$ resonance, then the currently observed 
minimum and maximum values of $\sin{(i)}$ of family members are 0.0742 and 
0.0782, respectively.  Neglecting possible changes in $\sin{(i)}$ after 
the family formation, which is motivated by the
fact that the local dynamics does not seems to particularly affect
asteroids in this region (see Fig.~\ref{fig: map_ai_cpv}),
these values set constraints on the possible terminal
ejection velocity parameter $V_{EJ}$ with which the family was created.
We generated synthetic families for values of $V_{EJ}$ from 5~m/s up to 
40 m/s.  Fig.~\ref{fig: inc_constr} 
show an $(a,\sin{(i)})$ projection of the initial orbital dispersion of 
the members of the family generated for $V_{EJ} = 15$~m/s (panel A) and 
$V_{EJ} = 25$~m/s.

For $V_{EJ} = 15$ m/s 6 particles (0.3\% of the total) had values of 
$\sin{(i)}$ outside the range of values currently observed and
$a > 2.995$~au, while for $V_{EJ} = 25$ m/s these number was 48 
(2.7\% of the total).  Based on these considerations, it seems unlikely
that the ejection velocity parameter
$V_{EJ}$ was larger than $25$~m/s, or a larger number of asteroids outside
the Hoffmeister family at $a > 2.795$~au would be visible today. 

\section{Ejection velocities evolution}
\label{sec: term_vel}

As previously discussed, the Hoffmeister family is one of the seven
families identified in \citet{Carruba_2016} as being characterized
by having a very leptokurtic distribution in the orthogonal
component $v_W$ of the currently estimated terminal ejection velocities,
and, therefore, of the asteroid inclinations.   As observed
in Sect.~\ref{sec: fam_ide}, this is mainly caused
by the interaction of this family with the ${\nu}_{1C}$ secular
resonance that strongly affects the distribution in proper inclination
of family members at lower values of semi-major axis, and, therefore,
their $v_W$ values.  Recently, \citet{Carruba_2016b} investigated how the
time evolution of $v_W$ values could be used to set constraints
on the initial values of the $V_{EJ}$ parameter for the Astrid family
Using the same approach, here we simulated fictitious Hoffmeister families
with the currently observed size-frequency distribution,
values of the parameters affecting
the strength of the Yarkovsky force typical of C-type asteroids according to 
\citet{Broz_2013}, i.e., bulk and surface density equal to 
${\rho}_{bulk}={\rho}_{surf} = 1300$~kg/m$^3$, thermal conductivity $K =0.01$~
W/m/K, thermal capacity equal to $C_{th} = 680$~J/kg/K, Bond albedo 
$A_{Bond} =0.02$ and infrared emissivity $\epsilon = 0.9$.
The fictitious families had values of $V_{EJ} = 10$, and
$20$~m/s, the most likely values of this parameter, according to the 
analysis of the previous section (values of $V_{EJ}$ larger or equal
to 25~m/s were deemed to be incompatible with the current
inclination distribution of the Hoffmeister family at large $a$).
Particles were integrated with
$SWIFT\_RMVSY$, the symplectic integrator developed by \citet{Broz_1999}
that simulates the diurnal and seasonal versions of the Yarkovsky effect, 
over 400 Myr and the gravitational influence of all major planets plus Ceres.
Values of $v_W$ were then obtained by inverting the third Gauss equation
\citep{Murray_1999}:

\begin{equation}
\delta i = \frac{(1-e^2)^{1/2}}{na} \frac{cos(\omega+f)}{1+e cos(f)} \delta v_W. 
\label{eq: gauss_3}
\end{equation}

where $\delta i= i-i_{ref}$, with $i_{ref}$ the inclination of the barycenter
of the family, and $f$ and $\omega+f$ assumed equal to 30$^{\circ}$ 
and 50.5$^{\circ}$, respectively. Results from \citet{Carruba_2016} show that 
the shape of the $v_W$ distribution is generally not strongly dependent
on the values of $f$ and $\omega+f$, except for values of $\omega+f$ close
to $\pm 90^{\circ}$ (which does not seems to be the case for the Hoffmeister
family, since these values would have produced a very
small inclination distribution).

\begin{figure*}
  \centering
  \begin{minipage}[c]{0.47\textwidth}
    \centering \includegraphics[width=2.8in]{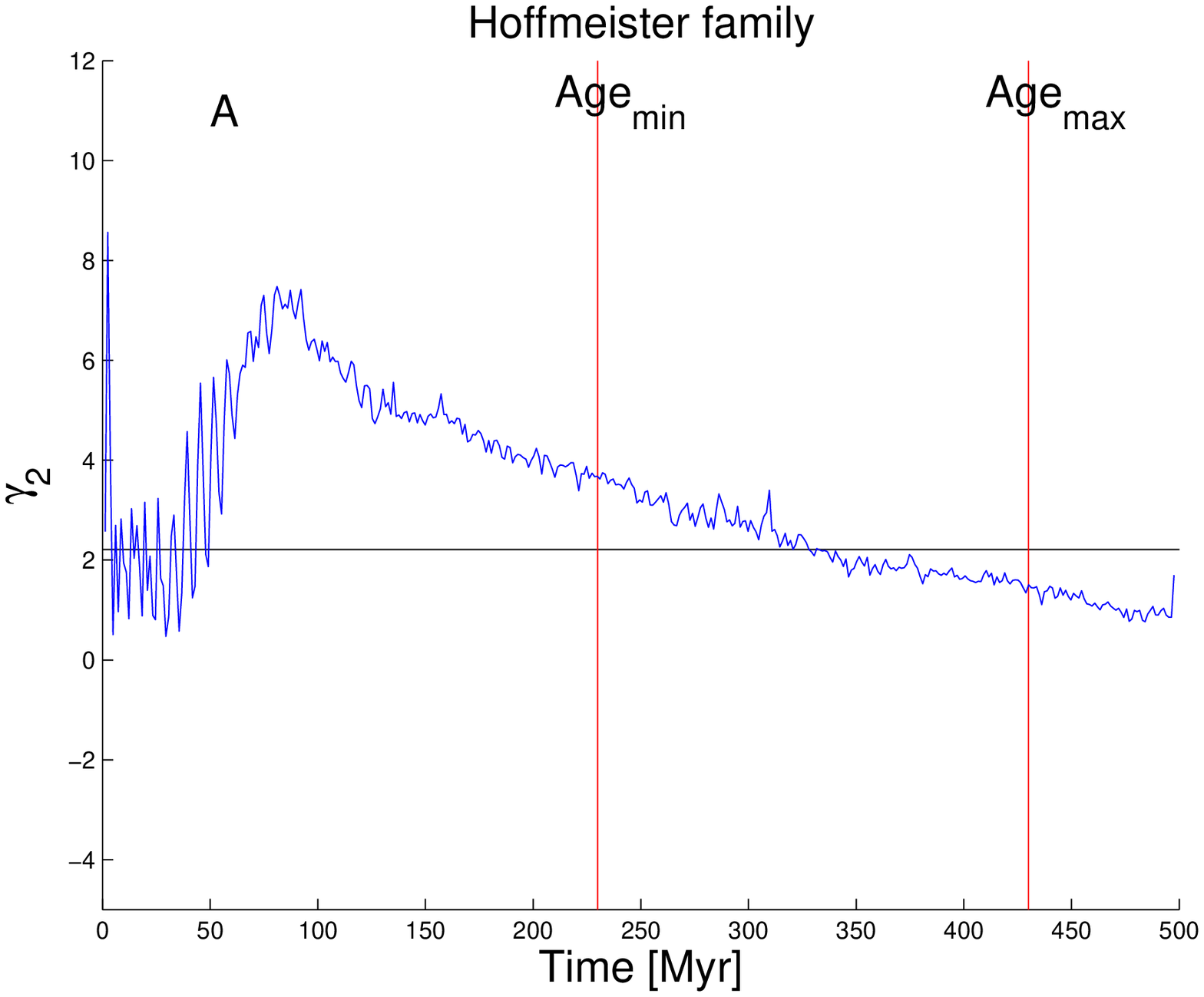}
  \end{minipage}%
  \begin{minipage}[c]{0.47\textwidth}
    \centering \includegraphics[width=2.8in]{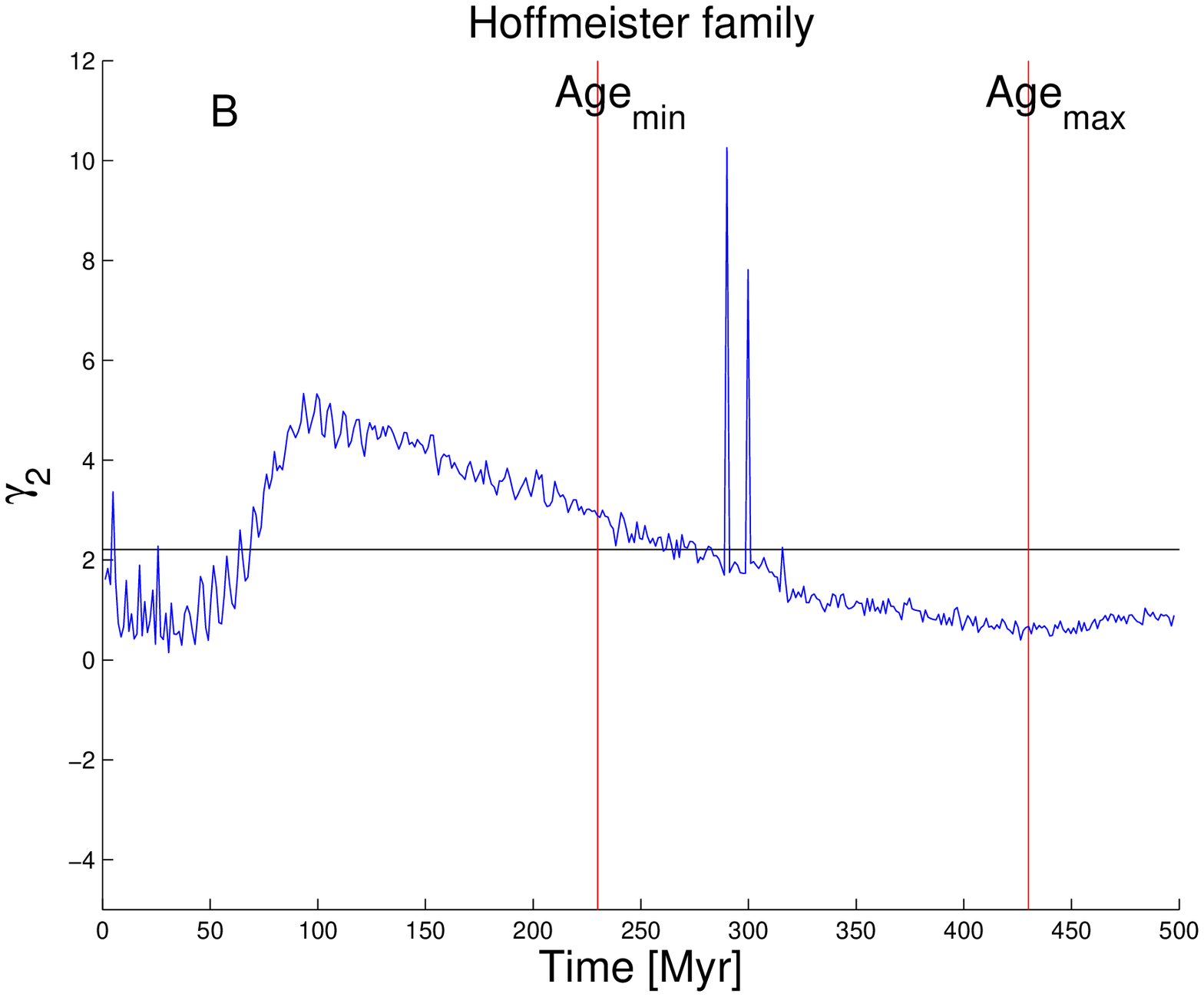}
  \end{minipage}
\caption{Time evolution of the Kurtosis parameter ${\gamma}_2(v_W)$ for members
  of a fictitious family with $V_{EJ} = 10$ (Panel A) and $20$ (panel B) m/s.
  The horizontal line identifies the current value of ${\gamma}_2(v_W)$
  for the Hoffmeister family (2.21).}
\label{fig: gamma2}
\end{figure*}

Our results are displayed in Fig.~\ref{fig: gamma2} for members
of a fictitious family with $V_{EJ} = 10$ (Panel A) and $20$~m/s (panel B).
Vertical lines display the maximum range of uncertainty for the family
age, according to \citet{Spoto_2015}, while the horizontal line shows
current value of ${\gamma}_2(v_W) = 2.21$ for the \citet{Nesvorny_2015}
Hoffmeister family (the ${\gamma}_2(v_W)$ for the
\citet{Milani_2014} family is quite similar and equal to 2.27).
In our computation of ${\gamma}_2(v_W)$ we neglected particles
that reached values of $\sin{(i)}$ beyond $\pm 4 \sigma({\sin{(i)}})$
from the family center.  Individual spikes in the time behaviour of
${\gamma}_2(v_W)$ in Fig.~\ref{fig: gamma2} are associated with single
particles that temporarily approached such extreme values of inclination. 
The present value of ${\gamma}_2(v_W)$ is first attained after 320 Myr for the
simulation with $V_{EJ} = 10$ m/s and after 260 Myr for the second
simulation.  It is lastly attained at 335 Myr in the first simulation
and at 280~Myr in the second, if we not consider fluctuations associated
with isolated spikes.  Overall, we have an upper limit for the Hoffmeister
age of 335~Myr for $V_{EJ} = 10$~m/s and of 280~Myr for $V_{EJ} = 20$~m/s.
In the next section we will try to further refine our family age estimate.
  
\section{Chronology}
\label{sec: chron}

\begin{figure*}

  \centering
  \begin{minipage}[c]{0.45\textwidth}
    \centering \includegraphics[width=3.in]{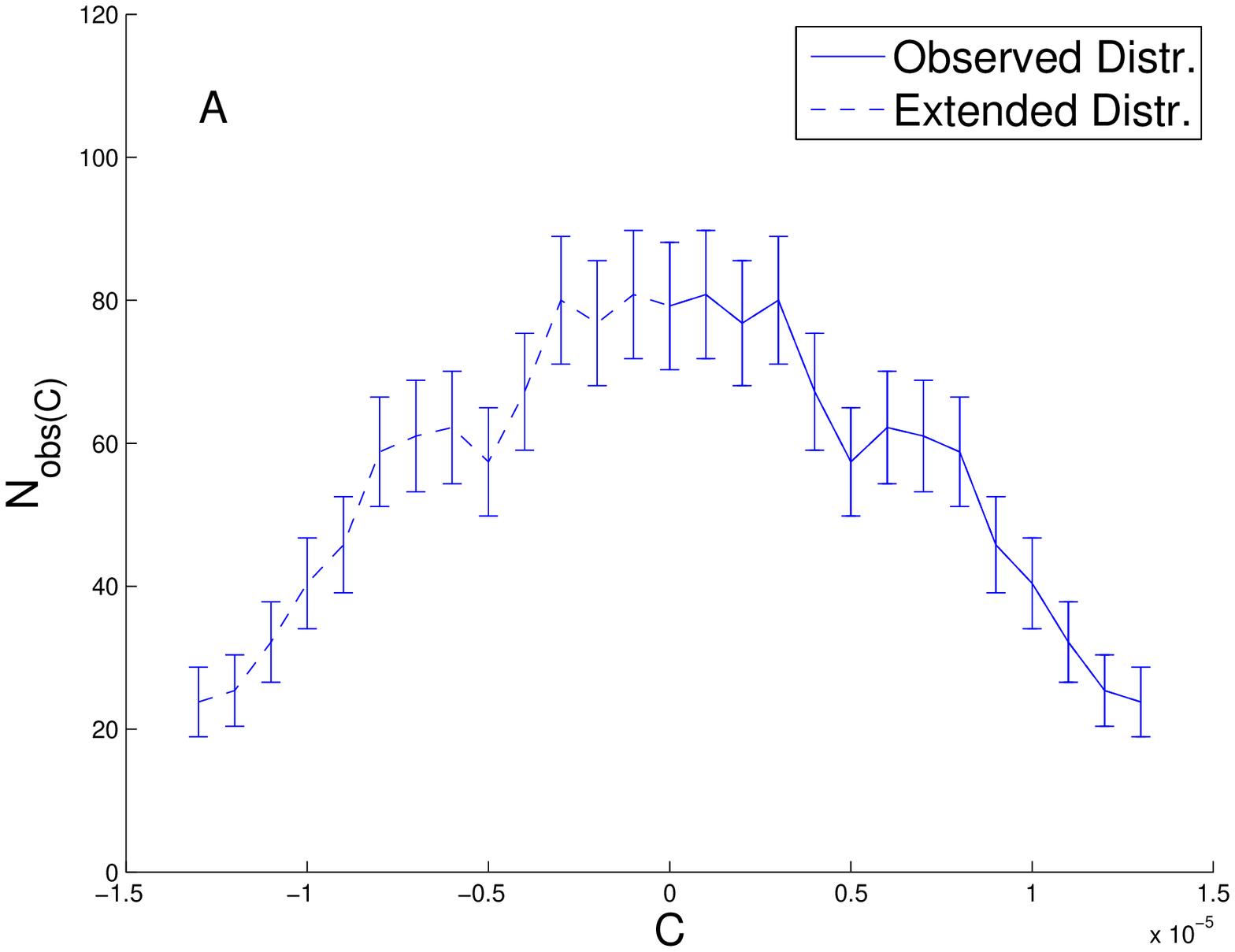}
  \end{minipage}%
  \begin{minipage}[c]{0.45\textwidth}
    \centering \includegraphics[width=3.in]{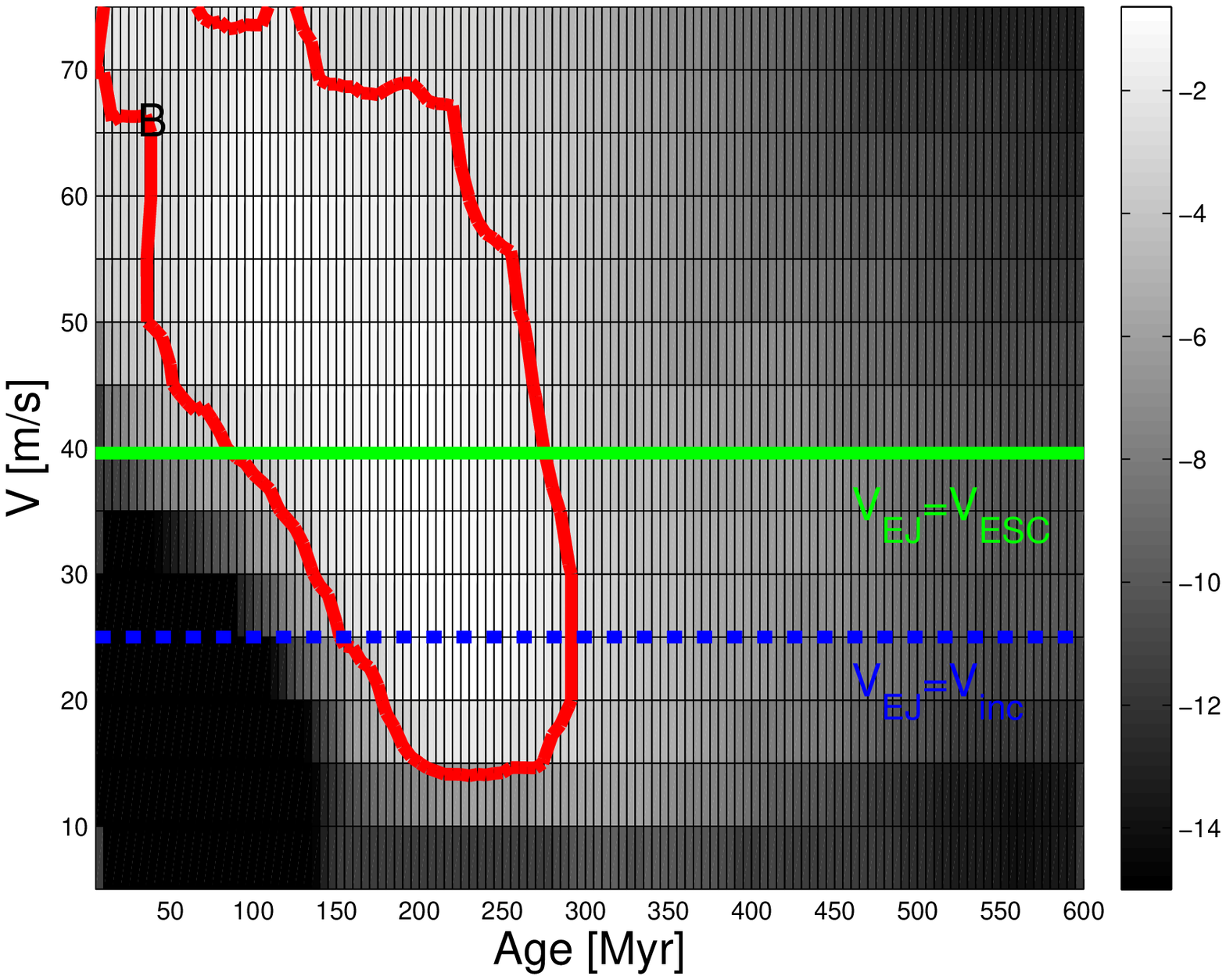}
  \end{minipage}

\caption{Panel A: Histogram of the distribution of $C$ values for the 
Hoffmeister family (blue line).  The dashed blue line displays the 
negative part of the $C$ distribution.
Panels B: target function ${\psi}_{\Delta C}$ values in ($Age,V_{EJ}$) 
plane for a symmetrical bi-modal distribution based on the 
$C$ negative values.  The horizontal green line display the value of 
the estimated escape velocity from the parent body, 
while the dashed blue line refers to the $V_{EJ} = 25$~m/s limit
obtained from the current inclination distribution in 
Sect.~\ref{sec: inc_constr}.  The red
lines display the contour level of ${\psi}_{\Delta C}$ associated
with a 1-sigma probability that the simulated and 
real distribution were compatible.}
\label{fig: Nobs_C_hoff}
\end{figure*}

Now that the analysis of the current inclination distribution and 
our ${\gamma}_2$ test provided independent constraint on the values 
of the $V_{EJ}$ parameter, we can try to obtain an independent age estimate
for this family.  We use the approach described in \citet{Carruba_2015a}
that employs a Monte Carlo method \citep{Vokrouhlicky_2006a,
Vokrouhlicky_2006b, Vokrouhlicky_2006c} to estimate the
age and terminal ejection velocities of the family members.
More details on the method can be found in \citet{Carruba_2015a}.  Essentially,
the semi-major axis distribution of simulated asteroid families is evolved
under the influence of the Yarkovsky effect (both diurnal and seasonal
version), the stochastic YORP force, and changes in values of the past
solar luminosity.  Distributions of a $C$-target function are then
obtained through the equation:

\begin{equation}
0.2H=log_{10}(\Delta a/C),
\label{eq: target_funct_C}
\end{equation}  

where $H$ is the asteroid absolute magnitude, and $\Delta a = a -a_{center}$
is the distance of each asteroid from its family center, here 
defined as the family center of mass.  For the Hoffmeister
family this is essentially equal to the semi-major axis of 1726
Hoffmeister itself.  We can then compare the
simulated $C$-distributions to the observed one by finding the minimum of a 
${\chi}^2$-like function:

\begin{equation}
{\psi}_{\Delta C}=\sum_{\Delta C}\frac{[N(C)-N_{obs}(C)]^2}{N_{obs}(C)},
\label{eq: psi}
\end{equation}

where $N(C)$ is the number of simulated objects in the $i-th$ $C$ interval,
and $N_{obs}(C)$ is the observed number in the same interval.    
Good values of the ${\psi}_{\Delta C}$ function are close to the number of 
the degrees of freedom of the ${\chi}^2$-like variable.  This is given
by the number of intervals in the $C$ minus the number of parameters 
estimated from the distribution (in our case, the family age and 
$V_{EJ}$ parameter).  Using only intervals with more than 10 asteroids,
to avoid the problems associated with small divisors in Eq.~\ref{eq: psi},
we have in our case 14 intervals for $C > 0$ 
(see Fig.~\ref{fig: Nobs_C_hoff}, panel A, results from the
\citet{Milani_2014} Hoffmeister family are compatible to within
the errors) and 2 estimated parameters, and, therefore, 12 degrees of freedom.
We only use positive values of $C$ so as to concentrate on the part
of the Hoffmeister family at larger semi-major axis, not affected
by the interaction with the ${\nu}_{1C}$ secular resonance.
If we assume that the ${\psi}_{\Delta C}$ probability 
distribution follows a law given by an incomplete gamma function of arguments 
${\psi}_{\Delta C}$ and the number of degrees of freedom, the value of 
${\psi}_{\Delta C}$ associated with a 3-sigma probability 
(or 99.7\%) of the simulated and real distributions being compatible is 
equal ${\psi}_{\Delta C}=2.73$ \citep{Press_2001}.

Results of our Monte Carlo simulation are shown in Fig.~\ref{fig: Nobs_C_hoff},
panel B.  If we neglect values of $V_{EJ} > 25$~m/s, that are not
likely to have occurred based on the current inclination distribution
of the part of the Hoffmeister family unaffected by the ${\nu}_{1C}$ secular
resonance, then our best-fit solution suggests an age of $220^{+70}_{-40}$~Myr
and an ejection parameter of $V_{EJ}= 20\pm5$~m/s.  Combining these
results with those of Sect.~\ref{sec: term_vel} (i.e., an age of less
than 280 Myr for $V_{EJ}= 20$~m/s), we can conclude
that the most likely age and ejection velocity parameter
of the Hoffmeister family are $220^{+60}_{-40}$~Myr and $V_{EJ}= 20\pm5$~m/s.

\section{Conclusions}
\label{sec: conc}

Our results could be summarized as follows:

\begin{itemize}

\item We first identified the Hoffmeister family in the domain of proper
  element \citep{Nesvorny_2015}, and used the results to define an
  orbital background region of this group.  Other families in the region
  are the Padua, Agnia, and Witt groups.  Of the 1819 members of the
  Hoffmeister dynamical group, 180 asteroids (9.9\% of the total)
  are in librating states of the ${\nu}_{1C}$ secular
  resonance, and while 54 members (3.0\% of the total)
  are in librating states of the $z_1 =g-g_6+s-s_6$.  Most of the Hoffmeister
  $z_1$ librators are objects at higher inclinations than those typical
  for the rest of the family, as also confirmed by our results
  for dynamical maps in the region (see Fig.~\ref{fig: map_ai_cpv}, panel A).

\item We revised the taxonomic and physical properties of the Hoffmeister
  family.  The Hoffmeister family is compatible with the break-up of
  a C-type object of low albedo (the mean geometric albedo of the family
  is 0.05).  Objects of low albedo ($0.025 < p_V < 0.075$) at lower
  inclinations than those of the Hoffmeister family may be former family
  members. No positive conclusion can be reached for low-$p_V$ objects
  at higher inclinations, because of possible contamination from the
  near X-type Padua family. 
  
\item We computed the fraction of particles that would reach regions
  in sine of inclination incompatible with the current distribution
  of Hoffmeister members that are not affected by the ${\nu}_{1C}$
  secular resonance, for fictitious families with different
  values of the $V_{EJ}$ ejection velocity parameter describing the standard
  deviation of their ejection velocity field, assumed to be
  Gaussian.  The current inclination distribution of the Hoffmeister
  family suggests that $V_{EJ}$ should be less than 25~m/s, or a
  larger fraction of its members would be currently observed at
  higher and lower inclinations.
  
\item We studied the dynamical evolution of two fictitious
  Hoffmeister families under the influence of the Yarkovsky
  effect for two values of the $V_{EJ}$ ejection velocity
  parameter, 10~m/s and 20~m/s.  Contrary to the case of the Astrid family,
  values of thermal conductivity ($K =0.01 W/m/K$) and mean density
  (${\rho}_{bulk} = {\rho}_{surface}= 1300~kg/m^3$) appropriate for
  a C-type class family such as Hoffmeister seems to produce
  results of ${\gamma}_2(v_W)$ evolution compatible
  with current estimates of the Hoffmeister family age.
  Current values of the ${\gamma}_2(v_W)$ parameter for the
  Hoffmeister family are reached until 335 Myr for $V_{EJ}= 10$~m/s
  and until 280 Myr for $V_{EJ}= 20$~m/s, which sets upper limits
  on the Hoffmeister family age.

\item We compute the age of the Hoffmeister family using a Monte Carlo
  approach for its Yarkovsky and YORP evolution.  Our best-fit
  solution, also accounting for the results from the ${\gamma}_2(v_W)$
  time behaviour analysis, suggests that the Hoffmeister family
  should be $220^{+60}_{-40}$~Myr old, with an ejection parameter
  $V_{EJ}= 20\pm5$~m/s.
  
\end{itemize}

Overall, an analysis of the ${\gamma}_2(v_W)$ time behaviour
provided invaluable constraints on the age and ejection velocity
field of a $v_W$ leptokurtic family, such as Hoffmeister, showing
once again, in our opinion, the importance that constraints
from secular dynamics (in this case the interaction of the Hoffmeister
family with the ${\nu}_{1C}$ secular resonance) could provide
in asteroid dynamics.

\section*{Acknowledgments}
We are grateful to the reviewer of this paper, Dr. Federica Spoto,
for comments and suggestions that greatly improved the quality of this
work.  We would also like to thank the S\~{a}o Paulo State Science Foundation 
(FAPESP) that supported this work via the grant 16/04476-8, and the
Brazilian National Research Council (CNPq, grant 305453/2011-4).
BN acknowledges support by the Ministry of Education, Science and
Technological Development of the Republic of Serbia, grant 176011.
This publication makes use of data products from the Wide-field 
Infrared Survey Explorer (WISE) and NEOWISE, which are a joint project 
of the University of California, Los Angeles, and the Jet Propulsion 
Laboratory/California Institute of Technology, funded by the National 
Aeronautics and Space Administration.

\bsp

\label{lastpage}

\end{document}